\documentclass{iopart}
\usepackage{amsfonts}
\usepackage{amssymb}
\usepackage{iopams}  
\usepackage{graphicx}% Include figure files
\newcommand{\bra}[1]{\langle #1|}
\newcommand{\ket}[1]{| #1 \rangle}

\begin{document}

\title[J.C. L\'{o}pez--Carre\~{n}o and H. Vinck--Posada]{Preparation of a three-photon state in a nonlinear cavity - quantum dot system}

\author{Juan Camilo L\'{o}pez-Carre\~{n}o}
\address{Grupo de \'{O}ptica e Informaci\'{o}n Cu\'{a}ntica, Departamento de F\'{i}sica, Universidad Nacional de Colombia, Bogot\'{a}, Colombia}
\ead{juclopezca@unal.edu.co}

\author{Herbert Vinck-Posada}
\address{Grupo de \'{O}ptica e Informaci\'{o}n Cu\'{a}ntica, Departamento de F\'{i}sica, Universidad Nacional de Colombia, Bogot\'{a}, Colombia}
\ead{hvinckp@unal.edu.co}

\begin{abstract}
We study theoretically the properties of a three-photon state prepared inside a semiconductor cavity, due to the interaction between a quantum dot and an electromagnetic field, and two consecutive spontaneous parametric downconversion (SPDC) processes. Thus, we consider a scheme involving three modes of the electromagnetic field, whose frequencies are given by the SPDC processes: $\omega_0 \rightarrow \omega_1 + \omega_2$, and $\omega_2 \rightarrow \omega_1 + \omega_1$. Furthermore, we study the low excitation regime, in which a three-photon state is accessible within the system's dynamics. 

\end{abstract}

%\submitto{\PS}
%\maketitle

\section{Introduction}

During the last few years, several research groups have been studying the light-matter interaction in quantum dots (QDs) embedded in semiconductor microcavities, both experimentally [1-6] and theoretically [7-15]. Such investigations lead to new phenomenology which in turn has lead to technological applications  [16-21].

%[1-6] -> \cite{Mexis-2001, Sergent-2012, Ates-2009, Yamamoto-2002, Vuckovic-2003, Pelton-2003}
%[7-15] -> \cite{Tejedor-2004, Vinck-2005, Yamamoto-2000, Exciton, Majumdar-2011, Laussy-2011, Luo-2012, Kaer-2013, Artuso-2013}
%[16-21] -> \cite{Loef-2009, Graham-2009, Solenov-2013, Carter-2013, Mashford-2013, Shoji-2013}

On the other hand, the generation of photons n-plets has been an interesting research branch, because it could allow researchers to prepare quantum states inside cavities which would be useful in quantum communication [22]. In particular, three-photons states can be obtained experimentally using Spontaneous Parametric down-conversion (SPDC) [23-28], and third-order optical non-linearities in assembled [29-31] systems. In this sense, although  several groups have managed to prepare and control specific quantum states \cite{Press-2008, Imamoglu-2012}, the preparation of arbitrary quantum states of light is still an experimental challenge.

%[22] -> \cite{Kiesel-2003}
%[23-28] -> \cite{Douady-2004, Corona-2011, Gravier-2008, Richard-2011, Dot-2012, Huebel-2010}
%[29-31] -> \cite{Persson-2004,Antonosyan,Rodrigo-2011}
%[32, 33] -> \cite{Press-2008, Imamoglu-2012}
%[34-37] -> \cite{Vamivikas-2004, Bravo-Abad-2007, Rodriguez-2007, Kuszelewicz-2012}

Bearing in mind the cavity Quantum Electrodynamics (cQED) description of the light-matter interactions and the preparations of quantum states of light via SPDC processes, we consider that it is possible to set up an experimental design in which the initial state inside a cavity can be prepared, and finely controlled. Even though the cavity does not have to be microscopic, the experimental design is scalable from those of semiconductor microcavities. This means that the problem and the obtained results are not restricted to the optical region of the electromagnetic spectrum. Therefore, in order to prepare the quantum state inside a cavity, a mesoscopic nonlinear crystal can be included to the experimental design. On the other hand, it has recently been demonstrated that Photonic Crystals (PhC) cavities are capable of enhancing the harmonic generation produced by either a $\chi^{(2)}$ or a $\chi^{(3)}$ nonlinearity, which would finally yield to a SPDC process [34-37].   Furthermore, it has been shown that the adequate geometry of the PhC \cite{Vamivikas-2004}, pump power \cite{Bravo-Abad-2007} and whether the cavity is singly or doubly resonant \cite{Rodriguez-2007} may lead to a $100\%$ conversion.

In this sense, we consider a semiconductor cavity in which there are a quantum dot (QD) and two nonlinear crystals. The former is coupled to a $\omega_0$ electromagnetic mode so the cavity is filled with $\omega_0$ photons. Afterwards, these photons go through the nonlinear crystals and two SPDC processes take place: $\omega_0 \rightarrow \omega_1 + \omega_2$, and $\omega_2 \rightarrow \omega_1 + \omega_1$. Taking into account an exciton pumping and $\omega_0$ photon leakage out of the cavity, which are both incoherent processes, a three-photon state is accessible in the $\omega_1$ electromagnetic mode. 

The nonlinear cavity-quantum dot system can be constructed using a GaAs substrate, over which several layers of $Al_x Ga_{1-x}As$, and a layer of $Al_y In_{1-y} As$ in which the quantum dots are localized, are grown using the Molecular Beam Epitaxy (MBE) method. In this particular construction, the optical properties are nonlinear \cite{Kuszelewicz-2012, Aguanno-2001}, and therefore could be used as basis for the experimental design of our system. Other such systems consists of PhC made periodically poled lithium niobate (PPLN) \cite{Huebel-2010, Myers-1995} or periodically poled potassium titanyl phosphate (KTP) \cite{Karlson-1997}, which give rise to the enhancement of the harmonic generation.  

A possible drawback of this kind of experimental designs lies in the fact that the inclusion  of QD into nonlinear cavities yield to physical phenomena such as the Kerr effect (due to the presence of other QDs in the cavity), the harmonic generation, or the Purcell effect (due to the spontaneous emission from a dipole source). Nevertheless, it has been shown that PhC cavities can lead to an enhancement of the nonlinear phenomena and supress the spontaneous emission via a photonic bang gap, which increases the $\chi^{(3)}$ nonlinearity \cite{Bravo-Abad-2007}. Furthermore, the SPDC processes can achieve a $100\%$ efficiency by using a doubly-resonant cavity \cite{Rodriguez-2007}, but producing such cavities is a challenge because they require confinement at two very different frequencies \cite{Bravo-Abad-2007}.

The rest of the paper is organized as follows. In Sec. 2, we present the model. In Sec. 3 we present our results and discuss its consequences. Finally, in Sec. 4 we provide an overview of the results and conclude.

\section{Model}

Our model considers the interaction between a QD and an electromagnetic mode inside a semiconductor cavity, followed by two consecutive SPDC processes. The latter lead to a total of three electromagnetic modes. The QD's elementary excitations \textit{-excitons}- are the result of an electron being promoted to the conduction from the valence band. In this paper, we model the interaction between the excitons and an electromagnetic mode with the usual Jaynes-Cummings model \cite{JCModel, JCModel1}, whose Hamiltonian is the following ($\hbar$ is taken as 1 along the paper),
\begin{equation}\label{JCHamiltonian}
H_{JC} = \omega_0 a_0^{\dagger}a_0 + \omega_{qd} \sigma^{\dagger}\sigma + g \left (a^{\dagger}_0\sigma + a_0\sigma^{\dagger}  \right ),
\end{equation}
where $a_0 (a_0^{\dagger})$ is the $\omega_0$ electromagnetic mode annihilation (creation) operator and $\sigma (\sigma^{\dagger})$ is the exciton annihilation (creation) operator. The $\omega_0$ electromagnetic mode and the QD's exciton are coupled with the interaction strength $g$ and their frequencies are close enough to resonance to allow for the rotating wave approximation; i.e. $\Delta = \omega_0 - \omega_{qd} \ll \omega_0,\, \omega_{qd}$  \cite{Carmichael}.

The two subsequent SPDC processes generate two more electromagnetic modes with frequencies $\omega_1$ and $\omega_2$. In the first process, one $\omega_0$ photon generates one $\omega_1$ and one $\omega_2$ photon $ \left ( \omega_0\rightarrow\omega_1 + \omega_2\right )$, whereas in the second process one $\omega_2$ generates two $\omega_1$ photons $ \left ( \omega_2\rightarrow\omega_1 + \omega_1\right )$. Both processes may be described in an effective way by the following Hamiltonian \cite{Antonosyan},
\begin{equation}
H_{SPDC} = \zeta \left (a_0 a_1^{\dagger} a_2^{\dagger} + a_0^{\dagger} a_1 a_2 \right )+ \xi \left (a_1^{\dagger 2} a_2 +  a_1^2 a_2^{\dagger} \right ),
\end{equation}
where $\zeta$ and $\xi$, are the rates at which the processes occur and the $a_i (a_i^{\dagger})$ are the $\omega_i$ mode annihilation (creation) operator. The physical system is depicted in the Fig. 1 and it is described by the Jaynes-Cummings plus SPDC Hamiltonians,
\begin{equation}\label{Hamiltonian}
H=H_{JC}+H_{SPDC}.
\end{equation}

The dynamical behaviour and the incoherent pumping and loss of the dot-cavity system is included in the Master equation, which in the Lindblad notation is written as,
\begin{equation}\label{Master eq.}
\dot{\rho}=i\left[\rho , H\right]+\frac{P}{2}\left(2\sigma^{\dagger}\rho\sigma - \lbrace \sigma\sigma^{\dagger},\rho\rbrace\right ) + \frac{\kappa}{2}\left(2a_0\rho a_0^{\dagger} -\lbrace a_0^{\dagger}a_0,\rho\rbrace\right),
\end{equation}
where $H$ is the Hamiltonian given in eq. (\ref{Hamiltonian}), $\kappa$ is the rate at which $\omega_0$ photons escape from the cavity and $P$ is the rate at which excitation is pumped to the cavity and is linked to the rate at which the electron-hole pairs relax into the dot. 

\begin{figure}[htb]
\centering \includegraphics[scale=0.3]{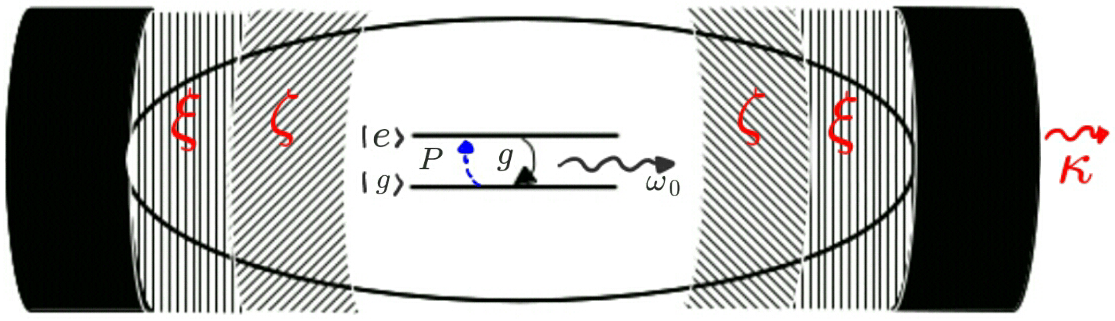} 
\caption{Diagram of the physical system. The nonlinear crystals ($\zeta$ and $\xi$) yield to the SPDC processes. }
\end{figure}

Furthermore, the system's energy levels and its connection via the master equation given in eq. (\ref{Master eq.}) is shown schematically in Fig. 2. Each energy level is associated to a quantum state written as $\ket{a,i,j,k}$, where $a$ is the QD state (either ground or excited), and $i$, $j$ and $k$ are the photon number in the $\omega_0$, $\omega_1$ and $\omega_2$ mode of the electromagnetic field, respectively. The presented scheme shows the energy levels accessible for the $\ket{g,n_0,n_1,n_2}$ state by just one process.

\begin{figure}[htb]
\centering \includegraphics[scale=0.43]{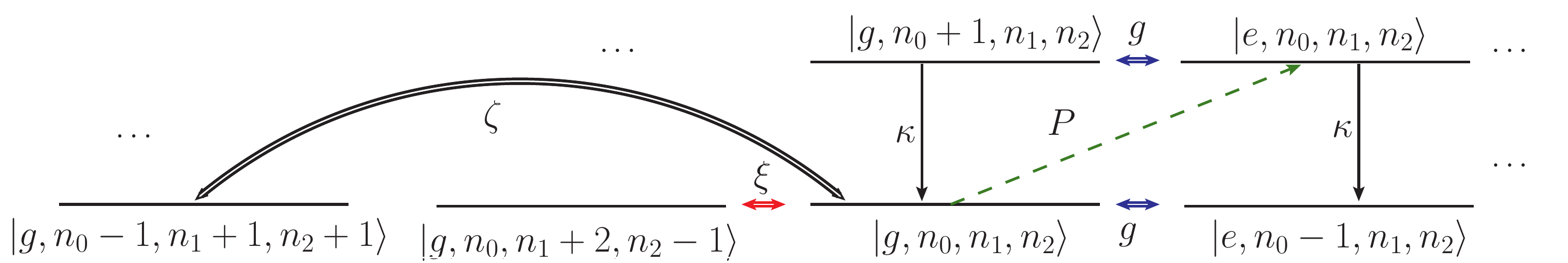} 
\caption{Ladder of energy levels for the QD-cavity system accessible from the $\ket{g,n_0,n_1,n_2}$ state by just one process. The energy levels are depicted by straight continuous lines corresponding to quantum states. The blue double lines correspond to the interaction between the QD and the $\omega_0$ mode. The black (red) double lines connect the accessible states via the SPDC process associated to $\zeta$ ($\xi$). The continuous black lines describe the $\omega_0$ leakage process, whereas the green piecewise line correspond to the incoherent exciton pumping.}
\end{figure}

\section{Results}

We solved the master equation given in (\ref{Master eq.}) numerically, using the following parameter values: he dipole-like interaction constant between the QD and the $\omega_0$-mode is set to be $g=5$ meV; the excitation energy of the QD is set as $\omega_{qd} = 500$ meV which in turn is tuned perfectly with the $\omega_0$-mode; i.e. $\omega_{qd} = \omega_0$. These values are the usual for $\lambda$ cavities operating in the infrared region of the electromagnetic spectrum. Furthermore, restraining our results to the low-excitation regime, we set the incoherent pumping rate $P$ to be $0$.$1\,\mu$eV, whereas the cavity losses for the fundamental mode $\omega_0$ is taken as $\kappa = 0$.$1$ meV. On the other hand, the SPDC rates are taken as $\xi =1$ meV and $\zeta = 3$ meV, following the recommendations presented in \cite{Antonosyan}. 

We consider the QD in its excited state and the electromagnetic field to be in a vacuum state in all of its modes, as the system's initial condition. With this set up, we observe that the state of the $\omega_1$ electromagnetic mode reaches a so-called three-photon state, which is a superposition of $3n$-photons Fock states, within the system's dynamics. This results are shown in the Fig. 3. 

In this way, we have obtained results similar to those presented in \cite{Antonosyan}, considering explicitly the interaction between a quantum dot in a semiconductor cavity and an electromagnetic field. This results are very interesting, since we have shown that a three-photon state can be prepared inside a semiconductor cavity made of Photonic Crystals capable of enhancing the harmonic generation produced by either a $\chi^{(2)}$ or a $\chi^{(3)}$ nonlinearity.

\begin{figure}[htb]
 \begin{minipage}{0.28\textwidth}
    \includegraphics[width=0.98\textwidth]{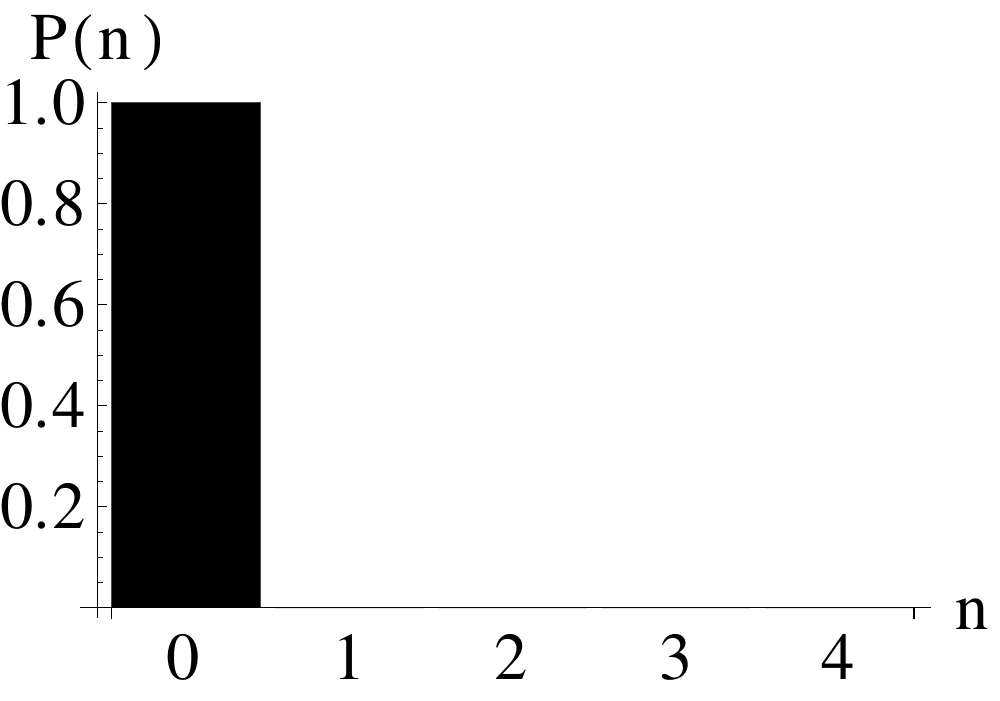}\\
    \includegraphics[width=0.98\textwidth]{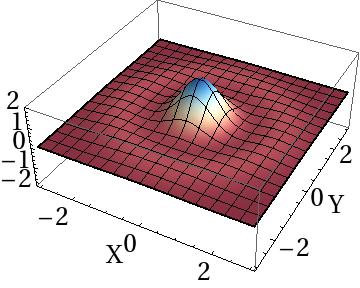}\\
    \includegraphics[width=0.98\textwidth]{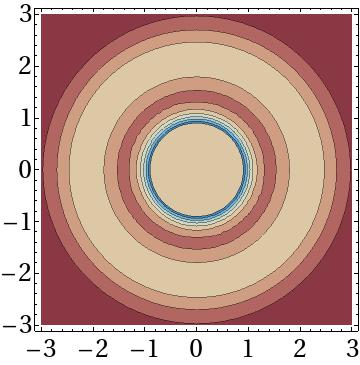}
    \begin{small}
    \begin{center}
    $t \kappa=0$
    \end{center}
    \end{small}
  \end{minipage}
  \ \hfill 
  \begin{minipage}{0.28\textwidth}
    \includegraphics[width=0.98\textwidth]{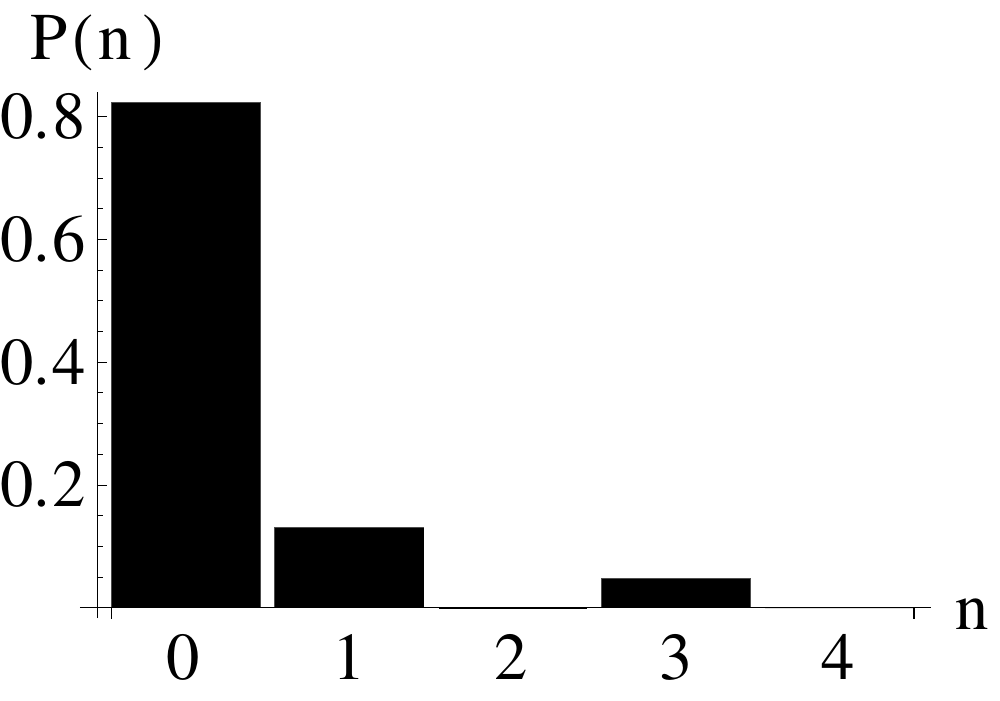}\\
    \includegraphics[width=0.98\textwidth]{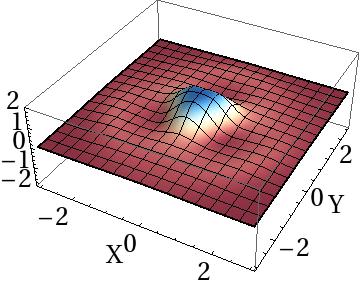}\\
    \includegraphics[width=0.98\textwidth]{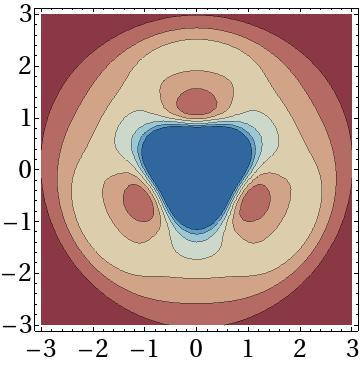}
    \begin{small}
    \begin{center}
    $t \kappa=0$.$216$
    \end{center}
    \end{small}
  \end{minipage}
  \ \hfill
  \begin{minipage}{0.28\textwidth}
    \includegraphics[width=0.98\textwidth]{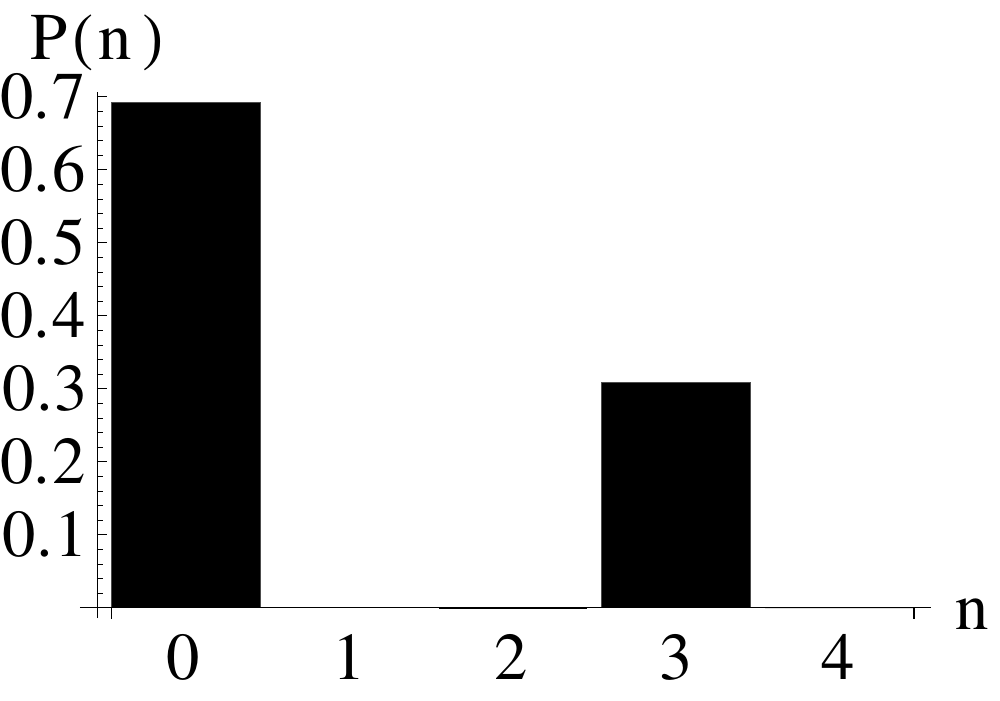}\\
    \includegraphics[width=0.98\textwidth]{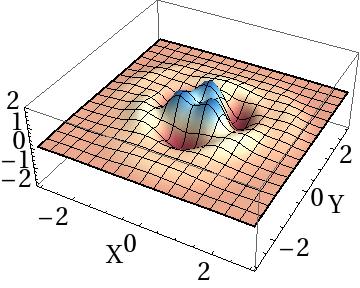}\\
    \includegraphics[width=0.98\textwidth]{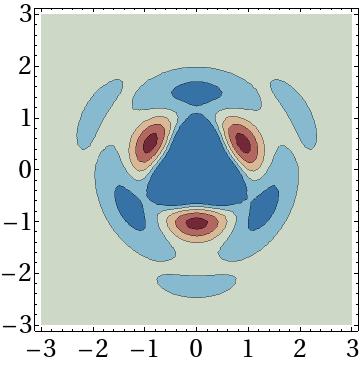}
    \begin{small}
    \begin{center}
    $t=0$.$328$
    \end{center}
    \end{small}
  \end{minipage}
\caption{(Color online). Photon number distribution (first row), Wigner functions (second row) and its contour plots (third row) for the $\omega_1$ mode for three different time intervals. The used parameters are $g/\kappa=50$, $\zeta/\kappa=30$, $\xi/\kappa=10$, $\kappa/P=1000$, $\omega_0=\omega_{qd}=500$ meV and $g=5$ meV.}
\end{figure}

\section{Summary and conclusions}

In this paper we have studied theoretically the preparation of a three-photon state inside a semiconductor cavity made of Photonic Crystals capable of enhancing the harmonic generation produced by either a $\chi^{(2)}$ or a $\chi^{(3)}$ nonlinearity. The three-photon state is the result of the interaction between a quantum dot embedded in the cavity and a $\omega_0$ mode of the electromagnetic field, and two spontaneous parametric downconversion processes yielding to two more modes of the electromagnetic field: $\omega_0 \rightarrow \omega_1 + \omega_2$, and $\omega_2 \rightarrow \omega_1 + \omega_1$. 
To study the system's dynamics we have solved a Lindblad master equation numerically considering both an incoherent excitation pump rate and $\omega_0$-photon leakage from the cavity. In this way, we have observed that the three-photon state is accessible within the dynamics in the $\omega_1$-mode in a low-excitation regime.

\newpage
\ack{This work has been supported by Colciencias within the project with code 110156933525, contract number 026-2013, and HERMES code 17432. Furthermore, we acknowledge the technical and computational support of the Grupo de \'{O}ptica e Informaci\'{o}n Cu\'{a}ntica.}

\appendix
\section{Dynamics of the density operator's matrix elements}

On the basis $\lbrace \ket{g,n_0,n_1,n_2};\ket{e,n_0,n_1,n_2}\rbrace$ of product states between the quantum dot and the $\omega_0$-, $\omega_1$-, and $\omega_2$-electromagnetic mode, the matrix elements of the density operator are,
\begin{equation}
\rho_{a,i,j,k;b,l,m,n} = \bra{a,i,j,k}\rho\ket{b,l,m,n},
\end{equation}
where $a$ and $b$ are either $g$ or $e$.

In this notation, the density operator's matrix elements satisfy the following differential equations:
\begin{scriptsize}
\begin{equation*}
\partial_t \rho_{g,i,j,k;g,l,m,n} = \left [i \omega_0 \left (l -i + \frac{m-j}{3} + 2\frac{n-k}{3}  \right )-\kappa \frac{l+i}{2}-P \right ] \rho_{g,i,j,k;g,l,m,n}
\end{equation*}
\begin{equation*}
+i g \left (\sqrt{l} \rho_{g,i,j,k;e,l-1,m,n} - \sqrt{i} \rho_{e,i-1,j,k;g,l,m,n}  \right ) + \kappa\sqrt{(i+1)(l+1)} \rho_{g,i+1,j,k;g,l+1,m,n}
\end{equation*}
\begin{equation*}
+ i \zeta \left ( \sqrt{l (m+1)(n+1)} \rho_{g,i,j,k;g,l-1,m+1,n+1} + \sqrt{(l-1)mn} \rho_{g,i,j,k;g,l+1,m-1,n-1} \right )
\end{equation*}
\begin{equation*}
-i \zeta \left ( \sqrt{(i+1)jk} \rho_{g,i+1,j-1,k-1;g,l,m,n}+ \sqrt{i(j+1)(k+1)} \rho_{g,i-1,j+1,k+1;g,l,m,n} \right )
\end{equation*}
\begin{equation*}
+ i \xi \left ( \sqrt{m(m-1)(n+1)} \rho_{g,i,j,k;g,l,m-2,n+1} + \sqrt{(m+1)(m+2)n} \rho_{g,i,j,k;g,l,m+2,n-1}  \right )
\end{equation*}
\begin{equation}
- i \xi \left ( \sqrt{(j+1)(j+2)k} \rho_{g,i,j+2,k-1;g,l,m,n} + \sqrt{j(j-1)(k+1)} \rho_{g,i,j-2,k+1;g,l,m,n}\right ),
\end{equation}
\end{scriptsize}

\begin{scriptsize}
\begin{equation*}
\partial_t \rho_{e,i,j,k;e,l,m,n} = \left [i \omega_0 \left (l -i + \frac{m-j}{3} + 2\frac{n-k}{3}  \right )-\kappa \frac{l+i}{2} \right ] \rho_{e,i,j,k;e,l,m,n}+ P \rho_{g,i,j,k;g,l,m,n}
\end{equation*}
\begin{equation*}
+i g \left (\sqrt{l+1} \rho_{e,i,j,k;g,l+1,m,n} - \sqrt{i+1} \rho_{g,i+1,j,k;e,l,m,n}  \right ) + \kappa\sqrt{(i+1)(l+1)} \rho_{e,i+1,j,k;e,l+1,m,n}
\end{equation*}
\begin{equation*}
+ i \zeta \left ( \sqrt{l (m+1)(n+1)} \rho_{e,i,j,k;e,l-1,m+1,n+1} + \sqrt{(l-1)mn} \rho_{e,i,j,k;e,l+1,m-1,n-1} \right )
\end{equation*}
\begin{equation*}
-i \zeta \left ( \sqrt{(i+1)jk} \rho_{e,i+1,j-1,k-1;e,l,m,n}+ \sqrt{i(j+1)(k+1)} \rho_{e,i-1,j+1,k+1;e,l,m,n} \right )
\end{equation*}
\begin{equation*}
+ i \xi \left ( \sqrt{m(m-1)(n+1)} \rho_{e,i,j,k;e,l,m-2,n+1} + \sqrt{(m+1)(m+2)n} \rho_{e,i,j,k;e,l,m+2,n-1}  \right )
\end{equation*}
\begin{equation}
- i \xi \left ( \sqrt{(j+1)(j+2)k} \rho_{e,i,j+2,k-1;e,l,m,n} + \sqrt{j(j-1)(k+1)} \rho_{e,i,j-2,k+1;e,l,m,n}\right ),
\end{equation}
\end{scriptsize}

\begin{scriptsize}

\begin{equation*}
\partial_t \rho_{g,i,j,k;e,l,m,n} = \left [i \omega_0 \left (l -i + \frac{m-j}{3} + 2\frac{n-k}{3}  \right ) +i\omega_{qd}-\kappa \frac{l+i}{2} - \frac{P}{2} \right ] \rho_{g,i,j,k;e,l,m,n}
\end{equation*}
\begin{equation*}
+i g \left (\sqrt{l+1} \rho_{g,i,j,k;g,l+1,m,n} - \sqrt{i} \rho_{e,i-1,j,k;e,l,m,n}  \right ) + \kappa\sqrt{(i+1)(l+1)} \rho_{g,i+1,j,k;e,l+1,m,n}
\end{equation*}
\begin{equation*}
+ i \zeta \left ( \sqrt{l (m+1)(n+1)} \rho_{g,i,j,k;e,l-1,m+1,n+1} + \sqrt{(l-1)mn} \rho_{g,i,j,k;e,l+1,m-1,n-1} \right )
\end{equation*}
\begin{equation*}
-i \zeta \left ( \sqrt{(i+1)jk} \rho_{g,i+1,j-1,k-1;e,l,m,n}+ \sqrt{i(j+1)(k+1)} \rho_{g,i-1,j+1,k+1;e,l,m,n} \right )
\end{equation*}
\begin{equation*}
+ i \xi \left ( \sqrt{m(m-1)(n+1)} \rho_{g,i,j,k;e,l,m-2,n+1} + \sqrt{(m+1)(m+2)n} \rho_{g,i,j,k;e,l,m+2,n-1}  \right )
\end{equation*}
\begin{equation}\label{Diff. eq. 1}
- i \xi \left ( \sqrt{(j+1)(j+2)k} \rho_{g,i,j+2,k-1;e,l,m,n} + \sqrt{j(j-1)(k+1)} \rho_{g,i,j-2,k+1;e,l,m,n}\right ),
\end{equation}
\end{scriptsize}
plus the hermitian conjugate of (\ref{Diff. eq. 1}).

Once the system of linear equations is solved, we obtain the density operator of the cavity-QD system as a function of time: $\rho (t)$. This operator has four quantum numbers associated, one to the QD and one to each of the modes of the electromagnetic field, and its matrix elements are thus given by,
\begin{equation}
\rho_{a,i,j,k;b,l,m,n}(t) = \bra{a,i,j,k}\rho(t)\ket{b,l,m,n}.
\end{equation}
Nevertheless, in this particular case we are only interested in the degree of freedom associated to the $\omega_1$-mode, so its convenient to consider the reduced (to the $\omega_1$ subsystem) density operator instead of the complete operator. The reduced operator is noted as $\rho^{(3)} (t)$, and is obtained from the complete operator by performing partial trace over all the remaining degrees of freedom:
\begin{equation}
\rho^{(3)}_{i,j} (t) = \bra{i}\rho^{(3)}(t)\ket{j} =  \sum_{a,n,m} \bra{a,n,i,m}\rho(t)\ket{a,n,j,m}.
\end{equation}

Finally, once we have obtained the reduced density operator for the $\omega_1$-mode, we compute its Wigner function as in, e.g. \cite{Barnett}.

\end{document}